\documentstyle[aps,psfig]{revtex}

\begin{document}

\title{Dispersion in Femtosecond
Entangled Two-Photon Interference}

\author{Jan Pe\v{r}ina, Jr.\footnote{On
leave from the Joint Laboratory of Optics of
Palack\'{y} University and Institute of Physics of Academy of Sciences of
the Czech Republic, 17. listopadu 50, 772 07 Olomouc,
Czech Republic.}\footnote{email: perina\underline{ }j@sloup.upol.cz},
Alexander V. Sergienko, \\
Bradley M. Jost, Bahaa E. A. Saleh,
Malvin C. Teich\footnote{email: teich@bu.edu}\\
Quantum Imaging Laboratory\footnote{URL: http://photon.bu.edu/%
teich/qil/QImaging.html} \\
Department of Electrical and Computer Engineering\\
           Boston University \\
           8 Saint Mary's Street,
           Boston, MA 02215,
           USA}

\maketitle

(short title: Entangled Two-Photon Interference)


\vspace{2mm}

\begin{abstract}
We theoretically investigate the quantum interference of entangled
two-photon states generated in a nonlinear crystal
pumped by femtosecond optical pulses.  Interference patterns
generated by the polarization analog of the Hong-Ou-Mandel
interferometer are studied.  Attention is devoted to the effects of the
pump-pulse profile (pulse duration and chirp) and the second-order
dispersion in both the nonlinear crystal and the interferometer's optical
elements. Dispersion causes the interference pattern to have an asymmetric
shape. Dispersion cancellation occurs in some cases.
\end{abstract}

\section{Introduction}
Significant consideration has
recently been given to the process of spontaneous parametric
down-conversion in nonlinear crystals pumped by
cw lasers \cite{MaWo,coinc,Klysh,Teich}.
The nonclassical properties
of entangled two-photon light generated by this process
have been used in many experimental schemes to elucidate
distinctions between the predictions of
classical and quantum physics \cite{quant}.
Coincidence-count measurements with entangled
two-photon states have revealed violations
of Bell's inequalities \cite{bell}, and have
been considered for use in nonclassical imaging \cite{Serg} and
quantum cryptography \cite{crypt}.

A new frontier in these efforts is the generation
of quantum states with
three correlated particles (GHZ states)
\cite{Zeil,three}, which would be most useful
for further tests of the predictions of quantum mechanics.
One way to create such states is to make use of pairs of
two-photon entangled states
that are synchronized in time, i.e.,
generated within a sharp time window \cite{three-two}. This
can be achieved by using femtosecond pump beams.
Also, successful quantum teleportation has already been
observed using femtosecond pumping \cite{teleport}.

For these reasons, the theoretical and experimental
properties of pulsed spontaneous parametric
down-conversion have been scrutinized \cite{Se,Ru,Gr,Ou}. It has been shown
that ultrashort pumping leads to a loss of visibility of the
coincidence-count interference pattern in type-II parametric
down-conversion \cite{Se,Ru,Gr}, and narrowband frequency filters
are required to restore the visibility \cite{three-two,Se,Gr}.

This paper is devoted to a theoretical investigation
of dispersion effects in femtosecond-pulsed spontaneous parametric
down-conversion. Particular attention is given to the effects of pump-pulse
chirp and second-order dispersion (in both the pump and down-converted
beams)
on the visibility and shape of the photon-coincidence pattern generated
by the polarization analog of the Hong-Ou-Mandel interferometer \cite{Mad}.
Dispersion cancellation, which has been extensively studied in the case of
cw pumping
\cite{Lar}, is also predicted to occur under certain conditions for
femtosecond down-converted pairs.

\section{Spontaneous parametric down-conversion
with an ultrashort pump pulse}

We consider a nonlinear crystal pumped by a strong coherent-state
field. Nonlinear interaction
then leads to the spontaneous generation of two
down-converted fields (the signal and the
idler) which are
mutually strongly correlated \cite{MaWo}. Such a correlation
can be conveniently described in terms
of the two-photon amplitude $ {\cal A}_{12} $ which is
defined as a matrix element of the
product of electric-field operators $ \hat{E}^{(+)}_1(z_1,t_1) $ and
$ \hat{E}^{(+)}_2(z_2,t_2) $
sandwiched between the entangled two-photon state $ |\psi^{(2)}\rangle $
(for details, see Appendix A) and
the vacuum state $ |{\rm vac} \rangle $:
\begin{equation}    
 {\cal A}_{12}(z_1,t_1,z_2,t_2)
   = \langle {\rm vac} |
  \hat{E}^{(+)}_1(z_1,t_1)  \hat{E}^{(+)}_2(z_2,t_2)
  |\psi^{(2)} (0,t)\rangle .
\end{equation}
The positive-frequency part $ \hat{E}^{(+)}_j $
of the electric-field
operator of the $ j $th beam is defined as
\begin{equation}      
 \hat{E}^{(+)}_j(z_j,t_j) = \sum_{k_j}
 e_j(k_j) f_j(\omega_{k_j}) \hat{a}_j(k_j) \exp(ik^v_j z_j - i
 \omega_{k_j} t_j )  , \hspace{1cm} j=1,2,
\end{equation}
where $ \hat{a}_{k_j} $ stands for the annihilation
operator of the mode with wave vector $ k_j $,
$ e_j(k_j) $ denotes
the normalization amplitude of the mode $ k_j $, and
$ f_j(\omega_{k_j}) $ characterizes an external
frequency filter placed in the $ j $th beam.
The symbols $ k^v_1 $ and $ k^v_2 $ denote
wave vectors in vacuum.

At the termination of the nonlinear interaction in the crystal,
the down-converted fields
evolve according to free-field evolution and thus
the two-photon amplitude $ {\cal A}_{12} $ depends only on the
differences $ t_1 - t $ and $ t_2 - t $.
When the down-converted beams propagate through
a dispersive material of the length $ l $, the entangled
two-photon state $ |\psi^{(2)} \rangle $
given in Eq. (A4) in Appendix A provides
the expression for $ {\cal A}_{12,l} $:
\begin{eqnarray}    
 {\cal A}_{12,l}(\tau_1,\tau_2) &=&
   C
 \int_{-L}^{0} dz \, \sum_{k_p} \sum_{k_1} f_1(\omega_{k_1})
 \sum_{k_2} f_2(\omega_{k_2})
 {\cal E}_p^{(+)}(0,\omega_{k_p}-\omega^0_p)
 \exp \left[ i (k_p-k_1-k_2) z  \right]
  \nonumber \\
  & & \mbox{} \times
 \exp\left[ i(\tilde{k}_1 + \tilde{k}_2) l \right]
 \delta( \omega_{k_p} - \omega_{k_1} - \omega_{k_2} )
 \exp [ -i\omega_{k_1}\tau_1 ]
 \exp [ -i\omega_{k_2}\tau_2 ] .
\end{eqnarray}
The times $ \tau_1 $ and $ \tau_2 $ are
given as follows:
\begin{equation}   
 -i \omega_{k_j}\tau_j = ik^v_j z_j - i \omega_{k_j}
 (t_j - t) , \hspace{1cm}  j=1,2  .
\end{equation}
The symbol $ {\cal E}_p^{(+)}(0,\omega_{k_p}-\omega^0_p) $ denotes
the positive-frequency
part of the envelope of the pump-beam electric-field amplitude
at the output plane of the crystal and $ \omega^0_p $ stands for
the central frequency of the pump beam;
the wave vectors $ k_p $, $ k_1 $, and $ k_2 $ ($ \tilde{k}_1 $
and $ \tilde{k}_2 $) are appropriate for the nonlinear
crystal (dispersive material). The symbol $ L $ means the length
of the crystal. The amplitudes $ e_1(k_1) $ and $ e_2(k_2) $ from
Eq. (2) are absorbed into the constant $ C $.

A typical experimental setup for coincidence-count
measurement is shown in Fig.~1.
\begin{figure}        
  \centerline{\hbox{\psfig{file=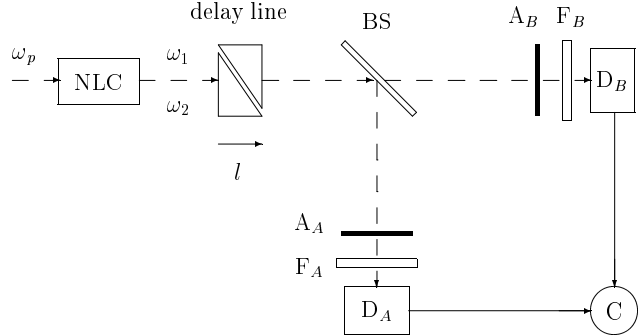,height=4.5cm}
     }}
  \vspace{3mm}
  \caption{Sketch of the system under consideration:
 a pump pulse at the frequency $ \omega_p $ generates
 down-converted photons at frequencies $ \omega_1 $ and $ \omega_2 $
 in the nonlinear crystal NLC. These waves
 propagate through a delay line of length $ l $
 and are detected at the detectors $ {\rm D}_{\rm A} $ and
 $ {\rm D}_{\rm B} $; BS denotes a beamsplitter;
 $ {\rm A}_{\rm A} $ and $ {\rm A}_{\rm B} $ are analyzers;
 $ {\rm F}_{\rm A} $ and $ {\rm F}_{\rm B} $ are frequency
 filters; and C indicates a coincidence device.}
\end{figure}
\noindent We consider
type-II parametric down-conversion for this exposition. In this case
two mutually perpendicularly polarized photons
are provided at the output plane of the crystal.
They propagate through a birefringent
material of a variable length $ l $ and then impinge on a 50/50
beamsplitter. Finally they are detected
at the detectors $ {\rm D}_A $ and $ {\rm D}_B $.
The coincidence-count rate $ R_c $ is measured
by a coincidence device C. The beams might be
filtered by the frequency filters $ {\rm F}_A $
and $ {\rm F}_B $ which can be placed in front of the detectors.
Analyzers rotated by 45 degrees with respect
to the ordinary and extraordinary axes of the nonlinear
crystal enable quantum interference between
two paths to be observed;
either a photon from beam 1 is detected
by the detector $ {\rm D}_A $ and a photon from beam 2 by
the detector $ {\rm D}_B $, or vice versa.

Including the effects of the beamsplitter and analyzers,
the coincidence-count rate $ R_c $ can be determined as follows
\cite{Se,Ru}:
\begin{equation}     
 R_c(l) = \frac{1}{4} \int_{-\infty}^{\infty} dt_A \,
 \int_{-\infty}^{\infty} dt_B \,
 \left| {\cal A}_{12,l}(t_A,t_B) -
 {\cal A}_{12,l}(t_B,t_A) \right|^2 ,
\end{equation}
where the two-photon amplitude $ {\cal A}_{12,l} $
is given in Eq. (3).

The normalized coincidence-count rate $ R_n $
is then expressed in the form:
\begin{equation}    
 R_n(l) = 1 - \rho(l) ,
\end{equation}
where
\begin{equation}      
 \rho(l) = \frac{1}{2R_0} \int_{-\infty}^{\infty} dt_A \,
 \int_{-\infty}^{\infty} dt_B \,
 {\rm Re} \left[ {\cal A}_{12,l}(t_A,t_B)
 {\cal A}^*_{12,l}(t_B,t_A) \right] ,
\end{equation}
and
\begin{equation}     
 R_0 = \frac{1}{2} \int_{-\infty}^{\infty} dt_A \,
 \int_{-\infty}^{\infty} dt_B \,
 \left| {\cal A}_{12,l}(t_A,t_B) \right|^2 .
\end{equation}
The symbol $ {\rm Re} $ denotes the real part of its argument.

\section{Specific models including second-order
dispersion}

Let us assume that
the nonlinear crystal and the optical material
in the path of the down-converted photons are both dispersive.
We proceed to generalize the models provided in Refs.
\cite{Se,Ru,Gr} by including the effects of second-order dispersion.

The wave vectors $ k_p(\omega_{k_p}) $, $ k_1(\omega_{k_1}) $,
and $ k_2(\omega_{k_2}) $
of the beams in the nonlinear crystal can be expressed
in the following form, when the effects of material dispersion
up to the second order are included \cite{SaTe}:
\begin{equation}     
 k_j(\omega_{k_j}) = k^0_j + \frac{1}{v_j} (\omega_{k_j} -
 \omega^0_j) + \frac{D_j}{4\pi} (\omega_{k_j} - \omega^0_j)^2 ,
 \hspace{1cm} j=p,1,2 .
\end{equation}
The inverse of group velocity $ 1/v_j $,
and the second-order dispersion coefficient $ D_j $,
are given by
\begin{eqnarray}    
 \frac{1}{v_j} &=& \frac{dk_j}{d\omega_{k_j}} \left.
 \right|_{\omega_{k_j}=\omega^0_j} , \\
 D_j &=& 2\pi \frac{d^2k_j}{d\omega_{k_j}^2} \left.
 \right|_{\omega_{k_j}=\omega^0_j} , \hspace{1cm} j=p,1,2 .
\end{eqnarray}
The symbol $ \omega^0_j $ denotes the central frequency of
beam $ j $.
The wave vector $ k^0_j $ is defined by the relation
$ k^0_j = k_j(\omega^0_j) $.

Similarly, the wave vectors $ \tilde{k}_1(\omega_{k_1}) $
and $ \tilde{k}_2(\omega_{k_2}) $ of the down-converted beams
in a dispersive material outside the crystal can be expressed as:
\begin{equation}    
 \tilde{k}_j(\omega_{k_j}) = \tilde{k}^0_j + \frac{1}{g_j}
 (\omega_{k_j} - \omega^0_j)
 + \frac{d_j}{4\pi} (\omega_{k_j} - \omega^0_j)^2 ,
 \hspace{1cm} j=1,2 ,
\end{equation}
where
\begin{eqnarray}    
 \frac{1}{g_j} &=& \frac{d\tilde{k}_j}{d\omega_{k_j}} \left.
 \right|_{\omega_{k_j}=\omega^0_j} ,  \\
 d_j &=& 2\pi \frac{d^2\tilde{k}_j}{d\omega_{k_j}^2} \left.
 \right|_{\omega_{k_j}=\omega^0_j} , \hspace{1cm} j=1,2,
\end{eqnarray}
and $ \tilde{k}^0_j = \tilde{k}_j(\omega^0_j) $.

We further assume that frequency filters with a Gaussian
profile, and centered around the central frequencies
$ \omega^0_1 $ and $ \omega^0_2 $, are incorporated:
\begin{equation}    
 f_j(\omega_{k_j}) = \exp\left[ - \frac{ (\omega_{k_j} -
 \omega^0_j)^2 }{ \sigma_j^2 } \right] ,  \hspace{1cm} j=1,2,
\end{equation}
where $ \sigma_j $ is the frequency width of the $ j $th
filter.

Assuming frequency- and wave-vector phase
matching for the central frequencies ($ \omega^0_p =
\omega^0_1 + \omega^0_2 $) and central wave vectors
($ k^0_p = k^0_1 + k^0_2 $), respectively,
the two-photon amplitude
$ {\cal A}_{12,l}(\tau_1,\tau_2) $ defined in Eq. (3) can be
expressed in the form:
\begin{eqnarray}    
 {\cal A}_{12,l}(\tau_1,\tau_2) &=& C_{\cal A}
 \exp(-i\omega^0_1\tau_1) \exp(-i\omega^0_2\tau_2)
 \int_{-L}^{0} dz \, \int d\Omega_p \,
 {\cal E}^{(+)}_p(0,\Omega_p)    \nonumber \\
  & & \mbox{} \times
 \int d\Omega_1 \, \exp \left[
 -\left( \frac{1}{\sigma_1^2} - i \frac{d_1 l}{4\pi}
 \right) \Omega_1^2 \right]
 \int d\Omega_2 \, \exp \left[
 -\left( \frac{1}{\sigma_2^2} - i \frac{d_2 l}{4\pi}
 \right) \Omega_2^2 \right] \delta(\Omega_p-\Omega_1-\Omega_2)
 \nonumber \\
 & & \mbox{} \times
 \exp\left[ i \left( \frac{\Omega_p}{v_p} - \frac{\Omega_1}{v_1}
 - \frac{\Omega_2}{v_2} \right) z \right]
 \exp\left[ i \left( \frac{D_p}{4\pi}\Omega_p^2 -
 \frac{D_1}{4\pi}\Omega_1^2 - \frac{D_2}{4\pi}\Omega_2^2 \right) z
 \right] \nonumber \\
 & & \mbox{} \times
 \exp\left[ -i\left( \tau_1 - \frac{l}{g_1} \right) \Omega_1 \right]
 \exp\left[ -i\left( \tau_2 - \frac{l}{g_2} \right) \Omega_2 \right]
 .
\end{eqnarray}
The frequencies $ \Omega_j $, $ \Omega_j = \omega_{k_j} -
\omega^0_j $, for $ j=1,2,p $
have been introduced in Eq. (16); $ C_{\cal A} $ denotes a constant.

We proceed to devote further attention to special cases.
We first consider an ultrashort pump
pulse with a Gaussian profile:
the envelope $ {\cal E}_p^{(+)}(0,t) $
of the pump pulse at the output plane of the crystal
then assumes the form \cite{Rudolph}:
\begin{equation}     
 {\cal E}_p^{(+)}(0,t) = \xi_{p0} \exp \left( - \frac{1+ia }{\tau_D^2}
 t^2 \right),
\end{equation}
where $ \xi_{p0} $ is the amplitude,
$ \tau_D $ is the pulse duration, and
the parameter $ a $ describes the chirp of the pulse.

The complex spectrum $ {\cal E}_p^{(+)}(z,\Omega_p) $ of the envelope
$ {\cal E}_p^{(+)}(z,t) $ is defined by
\begin{equation}     
 {\cal E}_p^{(+)}(z,\Omega_p) =  \frac{1}{2\pi}
 \int_{-\infty}^{\infty}
 dt \, {\cal E}_p^{(+)}(z,t) \exp( i \Omega_p t) .
\end{equation}

For a pulse of the form given in Eq. (17) we obtain:
\begin{equation}     
 {\cal E}_p^{(+)}(0,\Omega_p) = \xi_p \frac{\tau_D}{2\sqrt{\pi}
 \sqrt[4]{1+a^2} } \exp \left[ - \frac{ \tau_D^2}{
 4(1+a^2) } ( 1 - ia) \Omega_p^2 \right] ,
\end{equation}
where $ \xi_{p} = \xi_{p0} \exp[-i \arctan(a)/2] $.

Substituting Eq. (19) into Eq. (16) and using the identity
\begin{eqnarray}       
 \int_{-\infty}^{\infty} d\Omega_1 \,
 \int_{-\infty}^{\infty} d\Omega_2 \,
 \exp \left[ -\alpha_1 \Omega_1^2 - \alpha_2 \Omega_2^2
 -2\alpha_{12}\Omega_1\Omega_2 + i a_1\Omega_1 - i a_2\Omega_2
 \right]
 &=&
 \nonumber \\
  & & \hskip -10cm
  \frac{\pi}{ \sqrt{ \alpha_1\alpha_2 - \alpha_{12}^2  } }
 \exp \left[ - \frac{ a_1^2\alpha_2 + a_2^2\alpha_1 +
 2\alpha_{12}
 a_1a_2}{ 4( \alpha_1\alpha_2 - \alpha_{12}^2 ) } \right] ,
\end{eqnarray}
we arrive at the following expression for
the two-photon amplitude $ {\cal A}_{12,l}(\tau_1,\tau_2) $:
\begin{eqnarray}    
 {\cal A}_{12,l}(\tau_1,\tau_2) &=&
 C_{\cal A} \frac{\xi_p \tau_D}{
 2 \sqrt{\pi} \sqrt[4]{1+a^2} }
 \exp(-i\omega^0_1 \tau_1)
 \exp(-i\omega^0_2 \tau_2)
 A_{12,l}(\tau_1,\tau_2) , \\
 A_{12,l}(\tau_1,\tau_2) &=&
 \int_{-L}^{0} dz \,
 \frac{ 1}{ \sqrt{ \beta_1\beta_2 - \gamma^2 } }
 \exp \left[ - \frac{ c_1^2\beta_2 + c_2^2\beta_1 +
 2\gamma c_1 c_2 }{ 4( \beta_1\beta_2 - \gamma^2 ) }
 \right]  .
\end{eqnarray}
The functions $ \beta_j(z) $, $ c_j(z) $,
and $ \gamma(z) $ are defined as follows:
\begin{eqnarray}      
 \beta_j(z) &=& \frac{1}{\sigma_j^2} +
  b (1-ia) - i \frac{d_j}{4\pi} l
  - i \frac{D_p - D_j}{4\pi} z, \hspace{1cm}
 j=1,2  \nonumber \\
 c_j(z) &=& (-1)^{(j-1)} \left[ \left( \frac{1}{v_p} - \frac{1}{v_j}
 \right) z +  \frac{l}{g_j} - \tau_j \right] , \hspace{1cm} j=1,2
 \nonumber \\
 \gamma(z) &=& b (1-ia) - i \frac{D_p}{4\pi} z .
\end{eqnarray}
The parameter $ b $ is a characteristic parameter
of the pump pulse:
\begin{equation}      
 b = \frac{\tau_D^2}{4(1+a^2)} .
\end{equation}

The quantities $ \rho(l) $ and $ R_0 $ are then determined
in accordance with their definitions in Eqs. (7) and (8),
respectively.
The quantity $ \rho(l) $ as a function
of the length $ l $ of the birefringent material
then takes the form ($ \omega^0_1 = \omega^0_2 $ is assumed):
\begin{equation}    
 \rho(l) = \frac{ \pi^2 |C_{\cal A}|^2 |\xi_p|^2 \tau_D^2 }{ 2
 \sqrt{1+a^2} R_0 }  {\rm Re} \left\{
 \int_{-L}^{0} dz_1 \, \int_{-L}^{0} dz_2 \,
 \frac{ 1}{ \sqrt{ \bar{\beta}_1\bar{\beta}_2 - \bar{\gamma}^2 } }
 \exp \left[ - \frac{ \bar{c}_1^2\bar{\beta}_2 + \bar{c}_2^2
 \bar{\beta}_1 +
 2\bar{\gamma} \bar{c}_1 \bar{c}_2 }{ 4( \bar{\beta}_1\bar{\beta}_2
 - \bar{\gamma}^2 ) }
 \right] \right\} .
\end{equation}
The functions $ \bar{\beta}_j(z_1,z_2) $, $ \bar{c}_j(z_1,z_2) $,
and $ \bar{\gamma}(z_1,z_2) $ are expressed as follows:
\begin{eqnarray}      
 \bar{\beta}_j(z_1,z_2) &=& \frac{1}{\sigma_1^2} +
 \frac{1}{\sigma_2^2} - i \frac{d_j - d_{3-j}}{4\pi} l
 + 2 b - i \frac{D_p - D_j}{4\pi} z_1 +
 i \frac{D_p - D_{3-j}}{4\pi} z_2, \hspace{1cm}
 j=1,2  \nonumber \\
 \bar{c}_j(z_1,z_2) &=& \left( \frac{1}{v_p} - \frac{1}{v_1}
 \right) z_{j} - \left( \frac{1}{v_p} - \frac{1}{v_2}
\right) z_{3-j} + \left( \frac{1}{g_1} - \frac{1}{g_2}
 \right) l, \hspace{1cm} j=1,2
 \nonumber \\
 \bar{\gamma}(z_1,z_2) &=& 2 b - i \frac{D_p}{4\pi} (z_1-z_2) .
\end{eqnarray}

Similarly, the normalization constant $ R_0 $
is given by the expression:
\begin{equation}      
 R_0 = \frac{ \pi^2 |C_{\cal A}|^2 |\xi_p|^2 \tau_D^2 }{ 2
 \sqrt{1+a^2} }
 \int_{-L}^{0} dz_1 \, \int_{-L}^{0} dz_2 \,
 \frac{ 1}{ \sqrt{ \tilde{\beta}_1\tilde{\beta}_2 -
 \tilde{\gamma}^2 } }
 \exp \left[ - \frac{ \tilde{c}_1^2\tilde{\beta}_2 +
 \tilde{c}_2^2\tilde{\beta}_1 +
 2\tilde{\gamma} \tilde{c}_1 \tilde{c}_2 }{ 4(
 \tilde{\beta}_1\tilde{\beta}_2 -
 \tilde{\gamma}^2 ) }
 \right] ,
\end{equation}
where
\begin{eqnarray}      
 \tilde{\beta}_j(z_1,z_2) &=& \frac{2}{\sigma_j^2} +
 2 b - i \frac{D_p - D_j}{4\pi} (z_1 - z_2)
 , \hspace{1cm}
 j=1,2  \nonumber \\
 \tilde{c}_j(z_1,z_2) &=& \left( \frac{1}{v_p} - \frac{1}{v_j}
 \right) (z_{j} - z_{3-j}) , \hspace{1cm} j=1,2
 \nonumber \\
 \tilde{\gamma}(z_1,z_2) &=& 2 b - i \frac{D_p}{4\pi} (z_1-z_2) .
\end{eqnarray}

It is convenient to consider the pump pulse characteristics
at the output plane
of the crystal, i.e., to use the parameters $ \tau_D $ and $ a $.
They can be expressed in terms
of the parameters $ \tau_{Di} $ and $ a_i $ appropriate
for the input plane of the crystal:
\begin{eqnarray}    
 a &=& \left( \frac{\tau_{Di}^2 a_i}{4( 1+ a_i^2)} +
 \frac{D_pL}{4\pi} \right) \left(
 \frac{\tau_{Di}^2}{4(1+a_i^2)} \right)^{-1} , \nonumber \\
 \tau_D &=& \tau_{Di} \sqrt{ \frac{1+a^2}{1+ a_i^2} } .
\end{eqnarray}
In this case, the parameter $ b_i $
\begin{equation}      
 b_i = \frac{\tau_{Di}^2}{4(1+a_i^2)}
\end{equation}
has the same value as the parameter $ b $ defined in Eq. (24).

Ignoring second-order dispersion in all modes
($ D_p = D_1 = D_2 = 0 $), Eq.~(25) reduces to the following
analytical expression for the quantity $ \rho $:
\begin{equation}    
 \rho(\Delta\tau_l) = \sqrt{\frac{\pi}{2}}
 \frac{1}{|\Lambda| L} \frac{\tau_{Di}}{\sqrt{1+a_i^2}}
 {\rm erf} \left[ \frac{\sqrt{2}|\Lambda|}{D}
 \frac{\sqrt{1+a_i^2}}{\tau_{Di}} \left( \frac{DL}{2}
 - |\Delta\tau_l| \right) \right] ,
\end{equation}
in which
\begin{eqnarray}   
 D &=& \frac{1}{v_1} - \frac{1}{v_2} , \nonumber \\
 \Lambda &=& \frac{1}{v_p} - \frac{1}{2} \left(
 \frac{1}{v_1} + \frac{1}{v_2} \right) ,
\end{eqnarray}
and
\begin{equation}  
 \Delta \tau_l = \tau_l - DL/2  .
\end{equation}
The symbol $ {\rm erf} $ denotes the error function.
When deriving Eq. (31) the condition $ D > 0 $ was assumed.
In Eq. (33), $ \tau_l $ denotes
the relative time delay of
the down-converted beams in a birefringent material
of length $ l $ and is defined as follows:
\begin{equation}   
  \tau_l = \left( \frac{1}{g_2} - \frac{1}{g_1} \right) l .
\end{equation}

When second-order dispersion in the down-converted fields
is omitted, the interference pattern can be determined
for an arbitrary pump-pulse profile in terms of the autocorrelation
function of the pump pulse. For details, see Appendix B.

\section{Discussion}

We now proceed to examine the behavior of the normalized
coincidence-count
rate $ R_n $ on various parameters, from both analytical
and numerical points of view.

The profile of the interference dip in the coincidence-count
rate \cite{Mad} (described by $ \rho $ as a function of $ l $),
formed by the overlap of a pair of two-photon amplitudes,
can be understood as follows.
The expression in Eq. (7) for $ \rho(l) $ can be rewritten
in the form:
\begin{equation}    
 \rho(l) = \frac{1}{2R_0} \int_{-\infty}^{\infty} dt \,
 \int_{-\infty}^{\infty} d\tau \, \left[
 {\cal A}^r_{12,l}(t,\tau) {\cal A}^{r}_{12,l}(t,-\tau) +
 {\cal A}^i_{12,l}(t,\tau) {\cal A}^{i}_{12,l}(t,-\tau) \right] ,
\end{equation}
where
\begin{equation}    
 t = \frac{t_A + t_B}{2} , \hspace{1cm}
 \tau = t_A - t_B ,
\end{equation}
and
$ {\cal A}^r_{12,l} = \mbox{Re} [ {\cal A}_{12,l}] $;
$ {\cal A}^i_{12,l} = \mbox{Im} [ {\cal A}_{12,l}] $.
The symbol $ \mbox{Im} $ denotes
the imaginary part of the argument.
Hence, according to Eq. (35),
the overlaps of the real and imaginary
parts of the two-photon amplitudes
$ {\cal A}_{12,l}(t,\tau) $ and
$ {\cal A}_{12,l}(t,-\tau) $ determine the values
of the interference term $ \rho $.
The amplitude $ {\cal A}_{12,l}(t,-\tau) $ can be
considered as a mirror image of the amplitude
$ {\cal A}_{12,l}(t,\tau) $ with respect to
the plane $ \tau = 0 $.
When only
first-order dispersion in the optical material is taken
into account, the shape of the
two-photon amplitude $ {\cal A}_{12,l}(t,\tau) $
does not depend on the length $ l $; as $ l $ increases,
the amplitude $ {\cal A}_{12,l}(t,\tau) $
moves only in the $ t$-$\tau$ plane.
The shift in the $ \tau $-direction is important,
because it changes the degree of overlap of the
amplitudes. This reveals the origin of the
shape of the dip.

The overlap of the two-photon amplitudes can be interpreted
from the point-of-view of distinguishability of two paths leading
to coincidence detection \cite{Se}.
When the overlap is complete, the two paths cannot be
distinguished and the interference pattern has
maximum visibility. Incomplete overlap means
that the paths can be ``partially distinguished''
and thus the visibility is reduced.

We consider, in turn, the role played by pump-pulse duration
and chirp, second-order dispersion in the nonlinear down-converting
medium, second-order dispersion in the optical elements of
the interferometer, and dispersion cancellation.

\subsection{Pump-pulse duration and chirp}

In the absence of second-order dispersion and frequency filters,
a useful analytical expression for the two-photon
amplitude $ A_{12,l=0}(t,\tau) $
can be obtained:
\begin{equation}    
 A_{12,l=0}(t,\tau) =   \frac{ 4 \pi \sqrt{\pi}
 \sqrt[4]{1+a_i^2} }{
 \tau_{Di} |D| } \mbox{rect}\left( \frac{\tau}{DL} \right)
 \exp \left[ - \frac{ 1 + ia_i}{\tau_{Di}^2}
 \left( t + \frac{\Lambda}{D} \tau \right)^2 \right] .
\end{equation}
The coefficients $ D $ and $ \Lambda $ are defined
in Eq. (32). Equation (37) elucidates the role of pump-pulse
parameters as discussed below.
It is well known that for a cw-pump field the coincidence-count
rate $ R_n(\tau_l) $ forms a triangular
dip of width $ DL $ \cite{MaWo}.
The visibility is 100\%, indicating maximum
interference.
An ultrashort pump pulse of duration $ \tau_{Di} $ leads to a
loss of visibility (see Fig.\,2) but the width of the dip
remains unchanged \cite{Se}.
This can be understood from the shape of the two-photon amplitude
$ A_{12,l=0}(t,\tau) $ given in Eq.\,(37).
In the $ \tau $-direction the two-photon amplitude
is confined to the region $ 0 < \tau < DL $ for either cw
or an ultrashort pump pulse; this confinement is responsible for
the width of the dip.
The two-photon amplitude is confined in the $ t $-direction
by the ultrashort pump-pulse
duration
[see Eq. (37)]. The tilt (given by the ratio
$ \Lambda / D $, see Eq. (37))
of the amplitude
in the $ t$-$\tau $ plane leads to a loss of visibility
since the overlap
of the amplitudes $ A_{12,l}(t,\tau) $
and $ A_{12,l}(t,- \tau) $
for a given optimum value of $ l $ cannot be complete
for a nonzero tilt. The shorter
the pump-pulse duration,
the smaller the overlap, and the lower values of visibility
that result.
However, when values of the first-order dispersion
parameters are chosen such that $ \Lambda = 0 $,
the tilt is
zero [see Eq.\,(37)] and no loss of visibility occurs as the
pump-pulse duration shortens (for details, see \cite{Ru}).

\begin{figure}        
 \centerline{\hbox{\psfig{file=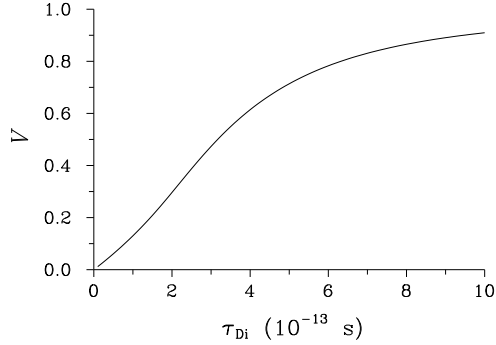,height= 4.5cm,angle=-90}
     }}
  \vspace{3mm}
  \caption{Visibility $ {\cal V} $ ($ {\cal V} = \rho /
 (2 - \rho ) $) as a function of the pump-pulse duration
 $ \tau_{Di} $;
 $ L = 3 $ mm, $ \sigma = \infty $ nm, and $ a_{i} = 0 $;
 values of the other parameters are zero.
In Figs. 2--8, the following parameters
apply. Values of the inverse
group velocities appropriate for the
BBO crystal with type-II interaction
at the pump wavelength $ \lambda_p = 397.5 $ nm,
and at down-conversion wavelengths
$ \lambda_1 = \lambda_2 = 795 $ nm are:
$ 1/v_p = 57.05 \times 10^{-13} $ s/mm,
$ 1/v_1 = 56.2 \times 10^{-13} $ s/mm, and
$ 1/v_2 = 54.26 \times 10^{-13} $ s/mm.
We assume that the optical materials for the
interferometer are quartz, for
which $ 1/g_1 = 51.81 \times 10^{-13} $ s/mm and
$ 1/g_2 = 52.08 \times 10^{-13} $ s/mm. }
\end{figure}

As indicated by the Eq.~(37) for the amplitude
$ A_{12,l=0} $, pump-pulse chirp (characterized by $ a_i $)
introduces a phase modulation of the two-photon amplitude in
the $ t $-direction. This modulation decreases the overall
overlap of the corresponding two-photon amplitudes, given as a sum
of the overlaps of their real and imaginary parts. Increasing values
of the chirp parameter $ a_i $ thus lead to a reduction of
visibility.
However, the width of the dip does not change. In fact,
it is the parameter $ b_i $ given in Eq. (30), combining
both the pulse duration $ \tau_{Di} $ and the
chirp parameter $ a_{i} $, that determines the visibility
in case of a Gaussian pump pulse. To be more specific the
parameter $ b_i $ is determined by the bandwidth
$ \Delta\Omega_p $ [$ \Delta\Omega_p = \sqrt{2} \sqrt{1+a_i^2} /
\tau_{Di} $, see Eq. (19)] of the pump pulse according to
the relation $ b_i = 1/ [2 (\Delta\Omega_p)^2] $.
Thus, more generally, it is the bandwidth of the pump pulse
that determines the interference pattern. As a consequence,
dispersion of the pump beam between the pump-pulse source and
the nonlinear crystal does not influence the interference pattern.

Examination of
Eqs. (B3) and (B4) in Appendix B shows that the dip remains symmetric since
the function $ \rho(\Delta\tau_l) $ in Eq. (B3) is
an odd function of $ \Delta\tau_l $ for an arbitrary
pump-pulse profile.

Frequency filters inserted into the down-converted
beams serve to broaden the two-photon amplitude
$ A_{12,l}(t,\tau) $ both in the $ t $- and
$ \tau $-direction. Broadening in the $ \tau $-direction
leads to wider dips, whereas that in the $ t $-direction
smooths out the effect of tilt discussed above and thereby
results in a higher visibility. The narrower the spectrum
of frequency filters, the wider the dip, and the higher
the observed visibility.
The effect of chirp is suppressed
by the presence of frequency filters,
because they effectively make the complex pump-pulse spectrum narrower
and hence diminish relative phase changes across such a
narrowed complex spectrum.

\subsection{Second-order dispersion
in the nonlinear crystal}

Second-order dispersion in the {\it pump beam}
causes changes in the pulse phase (chirp)
as the pulse propagates and this leads to broadening
of the pulse. The effect of such pump-pulse
broadening is transferred to the down-converted beams,
as is clearly shown by the behavior of the two-photon amplitude
$ A_{12,l}(t,\tau) $ illustrated in Fig.~3.
\begin{figure}        
 \centerline{\hbox{\psfig{file=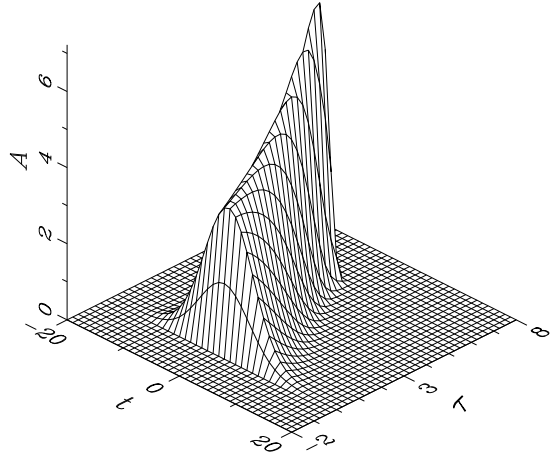,height= 6cm,angle=-90}
     }}
  \vspace{3mm}
  \caption{Absolute value of the two-photon amplitude
$ A = |A_{12,l=0}(t,\tau)|$
 for nonzero second-order dispersion of the pump beam; the variables $ t $
and $ \tau $ are in units
 of $ 10^{-13} $ s;
 $ \tau_{Di} = 1.55 \times 10^{-13} $ s,
 $ L = 3 $ mm, $ \sigma = 100 $ nm, $ D_p = 1 \times 10^{-25} $
 s$ {}^{2} $/mm, and
 $ a_{i} = 0 $; values of the other parameters are zero.}
\end{figure}
\noindent In this figure, the amplitude in the region near $ \tau = 0 $ s
has its origin near the output plane of the crystal
where the pump pulse is already broadened as a result
of its having propagated through the dispersive crystal.
At the other edge, near  $ \tau \approx 6 \times 10^{-13} $ s the
down-converted light arises from the
beginning of the crystal where the pump pulse has not
yet suffered dispersive broadening.
The profile of the interference dip is modified as follows:
An increase in the second-order
dispersion parameter $ D_p $
leads to an increase of visibility, but no change in the
width of the dip, as illustrated in Fig.\,4(a).
For appropriately chosen values of $ D_p $ a small local
peak emerges at the bottom of the dip [see Fig.\,4(a)].
Nonzero initial chirp ($ a_i $) of the pump beam
can provide a higher central peak but on the other hand
it reduces the visibility [see Fig.\,4(b)].
The peak remains, but is suppressed, in the presence of
narrow frequency filters.

\begin{figure}        
 \centerline{\hbox{(a) \hspace{3mm}
 \psfig{file=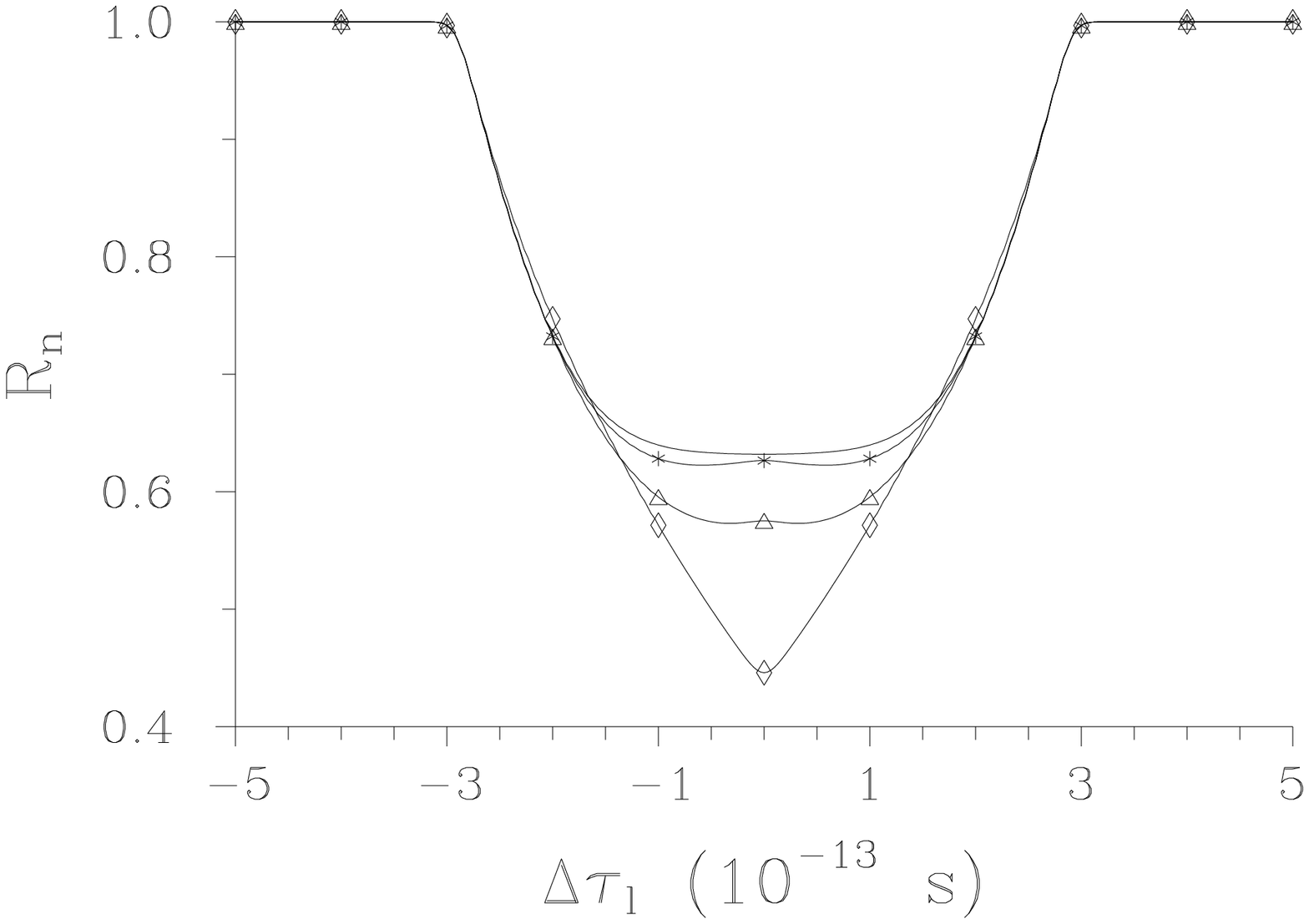,height=4.5cm,angle=-90}
 \hspace{10mm} (b) \hspace{3mm}
 \psfig{file=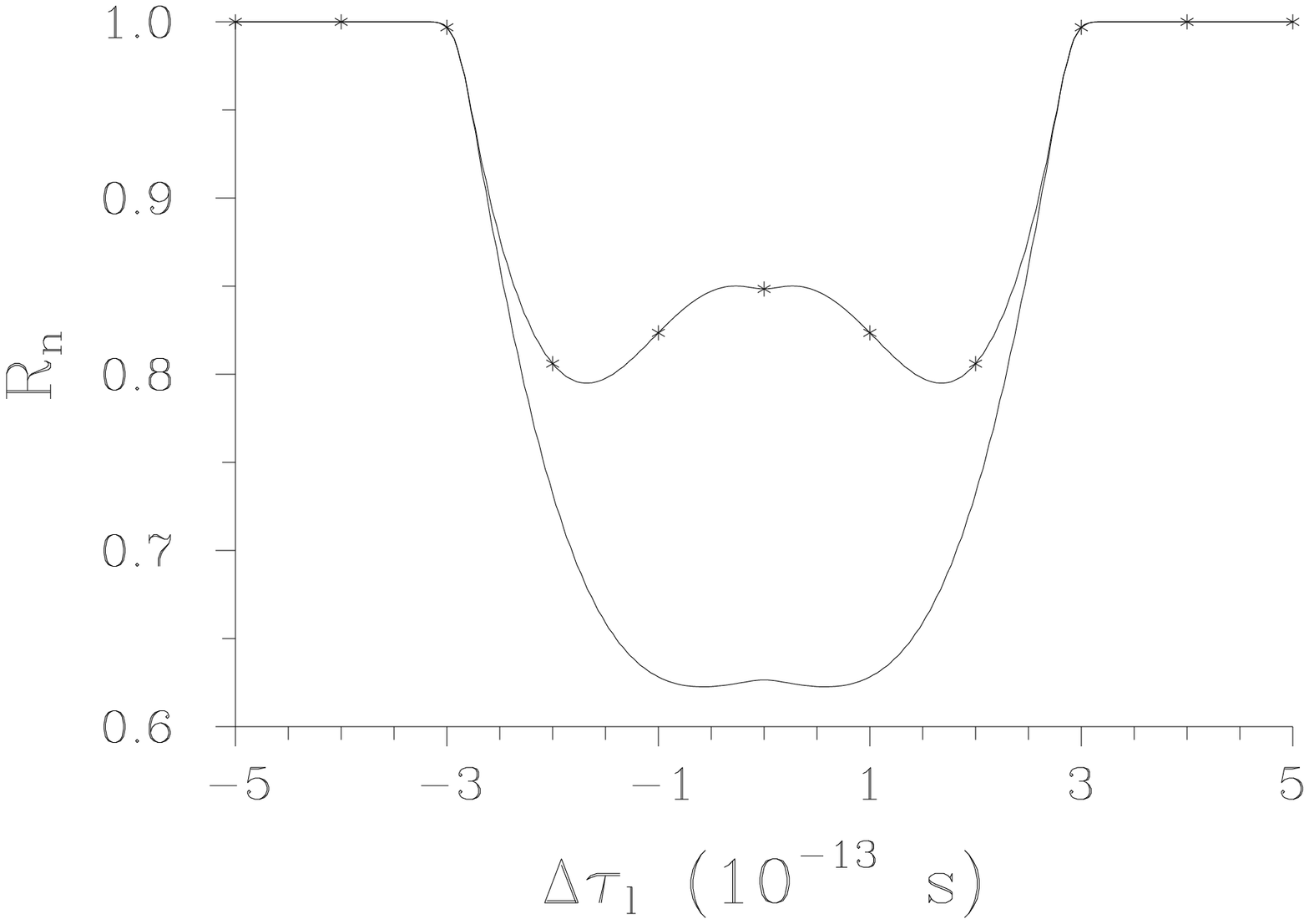,height= 4.5cm,angle=-90} }}
  \vspace{3mm}
  \caption{Coincidence-count rate $ R_n(\Delta\tau_l) $
 (a) for various values of the second-order dispersion
 parameter $ D_p $: $ D_p = 0 $
 s$ {}^{2} $/mm (plain curve),
 $ D_p = 5 \times 10^{-26} $
 s$ {}^{2} $/mm ($ \ast $), $ D_p = 1 \times 10^{-25} $
 s$ {}^{2} $/mm ($ \triangle $), and
 $ D_p = 3 \times 10^{-25} $
 s$ {}^{2} $/mm ($ \Diamond $), $ a_i = 0 $
 and
 (b) for various values of the chirp parameter $ a_i $:
 $ a_i = 0 $ (plain curve) and
 $ a_i = 2 $ ($ \ast $), $ D_p = 5 \times 10^{-26} $
 s$ {}^{2} $/mm;
 $ \tau_{Di} = 1.55 \times 10^{-13} $ s; $ L = 3 $ mm; $ \sigma =
 50 $ nm; values of the other parameters are zero. }
\end{figure}

Now we turn to second-order
dispersion in the {\it down-converted beams}
(nonzero $ D_1 $, $ D_2 $), which broadens the two-photon
amplitude $ A_{12,l}(t,\tau) $
in the $ \tau $-, as well as in the $t$-direction. As demonstrated
in Fig. 5, this leads
to a broadening of the dip, as well as asymmetry and oscillations
at its borders.
When values of $ D_1 $ increase,
visibility decreases at first and then later increases.
Nonzero chirp leads to a lower visibility, but
tends to suppress ocillations at the
borders of the dip. Frequency filters, which behave
as discussed above, suppress
asymmetry.

\begin{figure}        
 \centerline{\hbox{\psfig{file=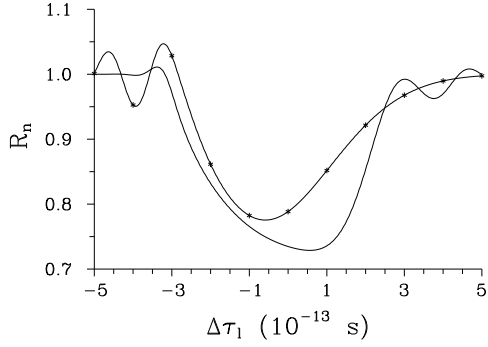,height= 4.5cm,angle=-90}
     }}
  \vspace{3mm}
  \caption{Coincidence-count rate $ R_n(\Delta\tau_l) $
 in case of second-order dispersion in beam
 1 (plain curve, $ D_1 = 1 \times 10^{-25} $
 s$ {}^{2} $/mm, $ D_2 = 0 $ s$ {}^{2} $/mm)
 and in beam 2 ($ \ast $,
 $ D_1 = 0 $ s$ {}^{2} $/mm, $ D_2 = 1 \times 10^{-25} $
 s$ {}^{2} $/mm);
 $ \tau_{Di} = 1.55 \times 10^{-13} $ s, $ L = 3 $ mm, and
 $ \sigma_1 = \sigma_2  = 50 $ nm;
 values of the other parameters are zero. }
\end{figure}

When second-order dispersion occurs in all three modes,
the two-photon amplitude $ A_{12,l}(t,\tau) $ is broadened
for smaller values of $ \tau $ (mainly owing to dispersion
in the pump beam) as well as for greater values of $ \tau $
(mainly owing to dispersion in the down-converted beams).
As a result, the interference pattern comprises all of the features
discussed above: a local peak may emerge at the bottom
of the dip, the dip is broadened and asymmetric, and
there occur oscillations at the borders of the dip.

To observe the above mentioned effects
caused by dispersion in a
nonlinear crystal, relatively large values of the
dispersion parameters $ D_p $, $ D_1 $, and
$ D_2 $ are required. For example, our simulations make use of parameter
values that are approximately an order of magnitude higher than
those of the
BBO crystals commonly used in type-II down-conversion-based interferometric
experiments.

\subsection{Second-order dispersion
in the interferometer's optical elements}

Second-order dispersion in an optical material
($ d_1 $, $ d_2 $) through
which down-converted photons propagate
leads to asymmetry of the dip. The dip is particularly
stretched to larger values of $ l $ (see Fig.~6) as a consequence
of the deformation and lengthtening of the two-photon amplitude
$ A_{12,l} $ in a dispersive material.
The higher the difference $ d_1 - d_2 $ of the
dispersion parameters, the higher the asymmetry and
the wider the dip; moreover its minimum is shifted
further to smaller values of $ l $ (see Fig.~6).
Asymmetry of the dip is also preserved
when relatively narrow frequency filters are used
though the narrowest filters remove it.
Chirp decreases visibility but the
shape of the dip remains unchanged.

\begin{figure}        
 \centerline{\hbox{\psfig{file=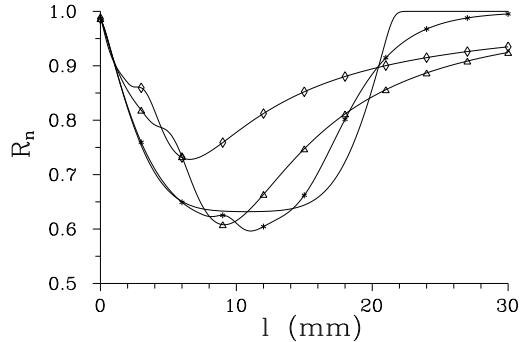,height= 4.5cm,angle=-90}
     }}
  \vspace{3mm}
  \caption{Coincidence-count rate $ R_n(l) $
 for various values of the second-order dispersion
 parameter $ d = d_1 - d_2 $ of an optical material;
 $ d = 0 $ s$ {}^{2} $/mm (plain curve), $ d = 1 \times 10^{-26} $
 s$ {}^{2} $/mm ($ \ast $),
 $ d = 5 \times 10^{-26} $ s$ {}^{2} $/mm ($ \triangle $), and
 $ d = 1 \times 10^{-25} $ s$ {}^{2} $/mm ($ \Diamond $);
 $ \tau_{Di} = 1.55 \times 10^{-13} $ s, $ L = 3 $ mm, and
 $ \sigma_1 = \sigma_2 = 50 $ nm; values of the other parameters
 are zero.}
\end{figure}

\subsection{Dispersion cancellation}

Asymmetry of the dip caused by second-order
dispersion in an optical material through which
down-converted photons propagate can be suppressed
in two cases. In the first case, for a pump pulse
of arbitrary duration, dispersion cancellation
occurs when the magnitude of second-order dispersion
in the path of the first photon (given by $ d_1 l $)
equals that of the second photon (given by $ d_2 l $).
This observation immediately follows from Eqs.\,(25) and (26), in which the
effect of second-order dispersion is prescribed by the parameter $
(d_1-d_2)l $. Dispersion cancellation is a result
of completely destructive interference between
the amplitudes $ A_{12,l}(t,\tau) $
and $ A_{12,l}(t,-\tau) $ for which there is nonzero
overlap. This is demonstrated in Fig.~7 for
$ l=25 $ mm, i.e. for which $ \rho = 0 $.

\begin{figure}        
 \centerline{\hbox{\psfig{file=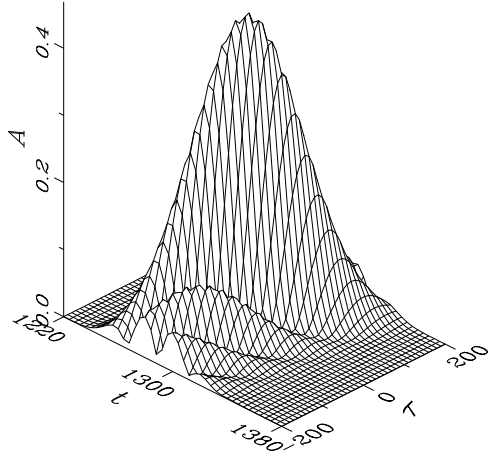,height= 6cm,angle=-90}
     }}
  \vspace{3mm}
  \caption{Absolute value
 of the two-photon amplitude $ A = |A_{12,l}(t,\tau)| $ for the same amount
 of second-order dispersion in the down-converted
 beams ($ d_1 = d_2 = 1 \times 10^{-25} $ s$ {}^{2} $/mm) for
 $ l = 25 $ mm; the variables $ t $ and $ \tau $ are
 in units of $ 10^{-13} $ s;
 $ \tau_{Di} = 1.55 \times 10^{-13} $ s,
 $ L = 3 $ mm, and $ \sigma = 100 $ nm;
 values of the other parameters are zero.}
\end{figure}

When the pulse duration is sufficiently long (in the cw regime)
dispersion cancellation occurs for arbitrary
magnitudes of second-order dispersion (given
by $ d_1 l $ and $ d_2 l $) present in the paths of
the down-converted photons.
The gradual suppression of the asymmetry of the dip as the pump pulse
duration increases is shown in Fig.~8.

\begin{figure}        
 \centerline{\hbox{\psfig{file=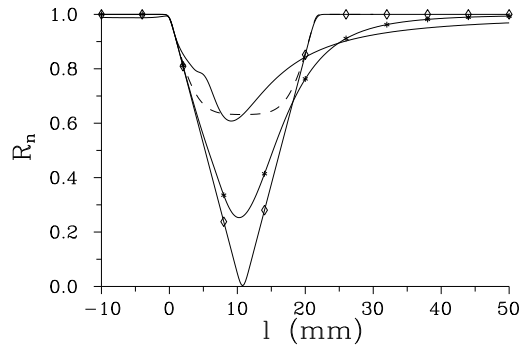,height= 4.5cm,angle=-90}
     }}
  \vspace{3mm}
  \caption{Coincidence-count rate $ R_n(l) $
 shows a gradual suppression of dispersion effects (asymmetry) as
 the pump-pulse duration increases; $ \tau_{Di} = 1.55 \times
 10^{-13} $ s
 (plain solid curve), $ \tau_{Di} = 5 \times 10^{-13} $ s
 ($ \ast $), and $ \tau_{Di} = 1 \times 10^{-11} $ s ($ \Diamond $),
 $ d = d_1 - d_2 = 5 \times 10^{-26} $ s$ {}^{2} $/mm; for comparison
 $ \tau_{Di} = 1.55 \times 10^{-13} $ s,
 $ d = 0 $ s$ {}^{2} $/mm (dashed curve);
 $ L = 3 $ mm; $ \sigma_1 = \sigma_2 = 50 $ nm;
 values of the other parameters are zero. }
\end{figure}

Dispersion cancellation has its
origin in the entanglement of the photons, i.e., in the
fact that the permitted values of the frequency $ \omega_1 $
and the frequency $ \omega_2 $ are governed
by the relation $ \delta(\omega_p - \omega_1
- \omega_2) $, where $ \omega_p $ lies within the pump-pulse
spectrum.

\section{Conclusion}

We have developed a description of two-photon type-II spontaneous parametric
down-conversion produced when ultrashort pulses from a femtosecond laser
are used to pump an appropriate nonlinear medium, as well as the associated
two-photon interference effects. The model includes frequency modulation of
the pump pulse (chirp) and dispersion in both the nonlinear crystal and
the interferometer's optical elements. The influence of these features on
the
depth and asymmetry characteristics of a photon-coincidence interference dip
have been established.

We showed that the interference pattern is determined by the
bandwidth of the pump pulse; the larger the bandwidth, the lower
the interference-pattern visibility. This implies that dispersion
of the pump beam before the nonlinear crystal does not influence
the interference pattern. Second-order dispersion
of the pump beam in the nonlinear crystal can result in the occurrence of a
local peak at the bottom of the interference dip. Second-order
dispersion of the down-converted photons in the crystal can result in
oscillations at the borders of the dip, whereas  dispersion of the
down-coverted photons in the interferometer's optical materials  (e.g., the
delay line) can produce an asymmetry in the dip. These effects can be used
to measure the dispersion parameters of both a nonlinear crystal
and an arbitrary optical material.
Dispersion cancellation has been revealed for pump pulses of arbitrary
duration when the amount of dispersion
in the two down-converted beams is identical and in general for sufficiently
long pump pulses.

\section*{Acknowledgments}

The authors thank J. Pe\v{r}ina and M. Atature for valuable
discussions.
This work was supported by the National Science Foundation
under Grant Nos. ECS-9800300 and ECS 9810355.
J. P. acknowledges support from Grant No. VS96028 of the
Czech Ministry of Education.

\appendix

\section{Determination of an entangled two-photon state}

The interaction Hamiltonian of the process of spontaneous
parametric down-conversion
can be written in the form \cite{MaWo}:
\begin{equation}    
 \hat{H}_{\rm int}(t) = \int_{-L}^{0} dz \, \chi^{(2)}
 E^{(+)}_p(z,t) \hat{E}^{(-)}_1(z,t) \hat{E}^{(-)}_2(z,t)
 + \mbox{h.c.} ,
\end{equation}
where $ \chi^{(2)} $ is the second-order susceptibility,
$ E^{(+)}_p $ denotes the positive-frequency part of
the electric-field amplitude of the pump field,
and $ E^{(-)}_1 $ ($ E^{(-)}_2 $)
is the negative-frequency part of the electric-field
operator of down-converted field 1 (2).
The nonlinear crystal extends from $ z=-L $ to $ z=0 $.
The symbol $ \mbox{h.c.} $ means Hermitian conjugate.

Expanding the interacting fields into harmonic
plane waves, the interaction Hamiltonian $ \hat{H}_{\rm int} $
in Eq. (A1) can be recast into the form:
\begin{eqnarray}     
 \hat{H}_{\rm int}(t) &=& C_{\rm int}
 \int_{-L}^{0} dz \, \sum_{k_p} \sum_{k_1}
 \sum_{k_2} \chi^{(2)}
 {\cal E}_p^{(+)}(0,\omega_{k_p}-\omega^0_p)
 \hat{a}_1^{\dagger}(k_1)
 \hat{a}_2^{\dagger}(k_2)
  \nonumber \\
 & &  \mbox{} \times \exp
 \left[ i (k_p-k_1-k_2)z - i (\omega_{k_p}-\omega_{k_1}-
 \omega_{k_2}) t \right]
 + \mbox{h.c.} ,
\end{eqnarray}
where $ C_{\rm int} $ is a constant.
The symbol $ {\cal E}_p^{(+)}(0,\omega_{k_p}-\omega^0_p) $ denotes
the positive-frequency
part of the envelope of the pump-beam electric-field amplitude
at the output plane
of the crystal; $ k_p $ stands for the wave vector of a mode
in the pump beam, and $ \omega^0_p $ stands for the central frequency
of the pump beam. The symbol $ \hat{a}_1^{\dagger}(k_1) $
($ \hat{a}_2^{\dagger}(k_2) $) represents the creation
operator of the mode with wave vector $ k_1 $ ($ k_2 $)
and frequency $ \omega_{k_1} $ ($ \omega_{k_2} $)
in the down-converted field 1 (2).
We note that the phases of all three interacting fields in space
are chosen in such a way that they are zero at the output
plane of the crystal.

The wave function $ |\psi^{(2)}(0,t)\rangle $ describing
an entangled two-photon state whose phases are set equal
to 0 at $ z=0 $ is given by:
\begin{equation}       
 |\psi^{(2)} (0,t)\rangle = \frac{-i}{\hbar}
 \int_{-\infty}^{t} dt'
 \hat{H}_{\rm int} (t') |{\rm vac} \rangle ,
\end{equation}
where $ |{\rm vac} \rangle $ denotes a multimode
vacuum state.

For times $ t $ sufficiently long so that the nonlinear interaction
is complete, the entangled two-photon state
$ |\psi^{(2)}(0,t) \rangle $ can be obtained in the
form:
\begin{eqnarray}      
 |\psi^{(2)}(0,t) \rangle &=&  C_{\psi}
 \int_{-L}^{0} dz \, \sum_{k_p} \sum_{k_1}
 \sum_{k_2}
 {\cal E}_p^{(+)}(0,\omega_{k_p}-\omega^0_p)
 \hat{a}_1^{\dagger}(k_1)
 \hat{a}_2^{\dagger}(k_2)
 \exp \left[ i(k_p-k_1- k_2) z  \right]
 \nonumber \\
  & & \mbox{} \times
 \delta( \omega_{k_p} - \omega_{k_1} - \omega_{k_2} )
 \exp\left[ i(\omega_{k_1} + \omega_{k_2}) t \right]
 |{\rm vac} \rangle .
\end{eqnarray}
The susceptibility $ \chi^{(2)} $
is included in the constant $ C_{\psi} $.
We note that for times during which the down-converted
fields are being created in the crystal, the appropriate
wave function differs from that in Eq. (A4).
However, detectors are placed at a sufficiently
large distance from the output plane of the crystal to
assure that such
``partially evolved'' states cannot be detected.

\section{Interference pattern for an arbitrary pump-pulse
profile}

We assume an arbitrary complex spectrum
$ {\cal E}^{(+)}_p(-L,\Omega_p) $ for the envelope of the
pump pulse at the input plane
of the crystal.
We further take into account the effect of second-order
dispersion only in the pump beam and
assume frequency filters of the
same width ($ \sigma_1 = \sigma_2 $). Under these conditions,
the normalized coincidence-count rate
$ R_n $ in Eq. (6) can be expressed in terms of the autocorrelation
function of the pump field.

Let us introduce the field $ {\cal E}^{(+)}_{p\sigma}(z, t) $
according to the definition:
\begin{eqnarray}      
 {\cal E}^{(+)}_{p\sigma}(z, t) =  \int_{-\infty}^{\infty}
 d\Omega_p \, {\cal E}^{(+)}_p(-L,\Omega_p) \exp\left[ i
 \frac{D_p(z+L)}{
 4\pi} \Omega_p^2 \right] \exp \left[ - \frac{\Omega_p^2}{\sigma^2}
 \right] \exp (-i\Omega_p t) ,
\end{eqnarray}
where $ \sigma = \sqrt{2} \sigma_1 $.
The above expression describes the propagation of the pump beam
through a dispersive material (a multiplicative term describing
first-order dispersion is not explicitly included here).
Equation (B1) also includes frequency filtering having its origin
in the filtering of the down-converted beams and
their entanglement with the pump beam.

The two-photon amplitude $ {\cal A}_{12,\tau_l}(\tau_1,\tau_2) $
can then be derived from the expression in Eq. (16):
\begin{eqnarray}     
 {\cal A}_{12,\tau_l}(\tau_1,\tau_2) &=& \frac{C_{\cal A}}{2}
 \exp(-i\omega^0_1 \tau_1 ) \exp(-i\omega^0_2 \tau_2)
  \nonumber \\
  & & \hskip -1.5cm \mbox{} \times
  \sqrt{\pi}\sigma \int_{-L}^{0} dz \,
 {\cal E}^{(+)}_{p\sigma}
 (z, (\tau_1 + \tau_l + \tau_2)/2  - \Lambda z)
 \exp \left[ - \frac{\sigma^2}{16} \left(
 \tau_1 + \tau_l - \tau_2 + D z \right)^2 \right] , \nonumber \\
\end{eqnarray}
where the parameters $ D $ and $ \Lambda $ are defined in Eq.
(32) and the relative time delay $ \tau_l $ of the down-converted
beams is introduced in Eq. (34).

The quantity $ \rho $ given in Eq. (7)
then has the form (again it is assumed that $ \omega^0_1 =
\omega^0_2 $):
\begin{eqnarray}      
 \rho(\Delta\tau_l) &=& \frac{ |C_{\cal A}|^2 \sqrt{2\pi} \pi
 \sigma}{4 R_0 } \nonumber \\
  & & \hskip -.5cm \mbox{} \times  {\rm Re} \left\{
 \int_{-L/2}^{L/2} dz_1 \, \int_{-L/2}^{L/2} dz_2 \,
 \gamma_{\sigma}(z_1,z_2,\Lambda(z_1-z_2))
 \exp \left[ - \frac{\sigma^2}{8} \left(
 \Delta\tau_l + \frac{D}{2}(z_1 + z_2) \right)^2
 \right]  \right\} , \nonumber \\
 & &
\end{eqnarray}
where $ \Delta \tau_l $ is defined in Eq. (33).
The correlation function $ \gamma_{\sigma}(z_1,z_2,x) $
of two pulsed fields at positions $ z_1 $ and $ z_2 $
is written as
\begin{equation}    
 \gamma_{\sigma}(z_1,z_2,x) = \int_{-\infty}^{\infty}
 dt \, {\cal E}^{(+)}_{p\sigma}(z_1 - L/2,t)
 {\cal E}^{(-)}_{p\sigma}(z_2 - L/2,t + x) .
\end{equation}
The constant $ R_0 $ occurring in Eq. (B3) is expressed
as follows:
\begin{eqnarray}      
 R_0 = \frac{ |C_{\cal A}|^2 \sqrt{2\pi} \pi
 \sigma}{4}
 \int_{-L/2}^{L/2} dz_1 \, \int_{-L/2}^{L/2} dz_2 \,
 \gamma_{\sigma}(z_1,z_2,\Lambda(z_1-z_2))
 \exp \left[ - \frac{\sigma^2 D^2}{32} (z_1 - z_2)^2
 \right] .
\end{eqnarray}

For a Gaussian pulse with the complex
spectrum as given in Eq. (19), the correlation function
$ \gamma_{\sigma} $ becomes
\begin{eqnarray}     
 \gamma_{\sigma}(z_1,z_2,\Lambda(z_1-z_2)) &=& \frac{\sqrt{\pi}
 \tau_{Di}^2}{ 2\sqrt{1+a_i^2} }  \frac{|\xi_{p}|^2}{\sqrt{
 \psi(z_1,z_2)}}
 \exp \left[ - \frac{\Lambda^2 (z_1-z_2)^2 }{4\psi(z_1,z_2)} \right],
 \nonumber \\
 \psi(z_1,z_2) &=& 2b_i + \frac{2}{\sigma^2} - i \frac{D_p}{4\pi}
 (z_1-z_2) ,
\end{eqnarray}
which, together with Eqs. (B3) and (B5), leads to
expressions which agree with those derived
from Eqs. (25) and (27).
The parameter $ b_i $ is defined in Eq. (30).

The experimental setup without frequency
filters ($ \sigma \rightarrow \infty $)
is of particular interest. In this case,
using the identity $ \sqrt{\pi} \sigma
\exp( - \sigma^2 y^2 / 4 ) \rightarrow 2\pi
\delta(y) $ for $ \sigma \rightarrow \infty $,
Eqs. (B3) and (B5) provide
a useful expression for the function $ \rho(\Delta\tau_l) $:
\begin{eqnarray}     
 \rho(\Delta\tau_l) &=& \frac{ 1 }{ \gamma_{\infty}(0,0,0) L}
 {\rm Re} \left\{ \int_{-L/2}^{L/2}
 dz \, {\rm rect} \left( z/L + 1/2 + 2\Delta\tau_l/(DL)
 \right) \right. \nonumber \\
 & & \mbox{} \times \left. \gamma_{\infty}(z, -z - 2\Delta\tau_l/D,
 2z + 2\Delta\tau_l/D ) \right\} ,
\end{eqnarray}
where $ {\rm rect}(x) $ is the rectangular function
($ {\rm rect} (x) = 1 $ for $ 0<x<1 $ and
$ {\rm rect} (x) = 0 $ otherwise).

\end{document}